\documentclass[twocolumn,aps,prb,showkeys,superscriptaddress,letterpaper,nolongbibliography]{revtex4-2}
\usepackage[utf8]{inputenc}
\usepackage{amsmath}
\usepackage{amssymb}
\usepackage{graphicx}
\usepackage[usenames,dvipsnames]{xcolor}
\usepackage{bm}% bold math
\renewcommand{\mathbf}{\bm}

\usepackage[unicode=true,
 bookmarks=true,bookmarksnumbered=false,bookmarksopen=false,
 breaklinks=false,pdfborder={0 0 0},pdfborderstyle={},backref=false,colorlinks=true,]
 {hyperref}
\hypersetup{linkcolor=blue,citecolor=blue,urlcolor=blue}

\DeclareMathOperator{\Real}{Re}\renewcommand{\Re}{\Real}
\DeclareMathOperator{\Imag}{Im}\renewcommand{\Im}{\Imag}

\begin{document}

%\title{Subgap states and quantum phase transitions in superconductors with extended magnetic scatterers}
\title{Subgap states and quantum phase transitions in one-dimensional superconductor-ferromagnetic insulator heterostructures}

\author{Javier Feijoo}%\email{javier.feijoo@fisica.unlp.edu.ar}

\affiliation{Instituto de Física La Plata - CONICET, Diag 113 y 64 (1900) La Plata, Argentina}
\affiliation{Departamento de Física, Universidad Nacional de La Plata, cc 67, 1900 La Plata, Argentina.}

\author{Aníbal Iucci}%\email{iucci@fisica.unlp.edu.ar}

\affiliation{Instituto de Física La Plata - CONICET, Diag 113 y 64 (1900) La Plata, Argentina}
\affiliation{Departamento de Física, Universidad Nacional de La Plata, cc 67, 1900 La Plata, Argentina.}

\author{Alejandro M. Lobos}%\email{alejandro.martin.lobos@gmail.com}

\affiliation{Instituto Interdisciplinario de Ciencias Básicas (CONICET-UNCuyo)}
\affiliation{Facultad de Ciencias Exactas y Naturales, Universidad Nacional de Cuyo, 5500 Mendoza, Argentina}

\begin{abstract}
We theoretically study the spectral properties of a one dimensional  semiconductor-superconductor-ferromagnetic insulator (SE-SU-FMI) hybrid nanostructure, motivated by recents experiments where such devices have been fabricated using epitaxial growing techniques.  
We model the hybrid structure as a one-dimensional single-channel semiconductor nanowire under the simultaneous effect of two proximity-induced interactions: superconducting pairing and a (spatially inhomogeneous) Zeeman exchange field. The coexistence of these competing interactions generates a rich quantum phase diagram and a complex subgap Andreev bound state (ABS) spectrum. By  exploiting the symmetries of the problem, we classify the solutions of the Bogoliubov-de Gennes equations into even and odd ABS with respect to the spatial inversion symmetry $x\to -x$. We find the ABS spectrum of the device as a function of the different parameters of the model: the length $L$ of the coexisting SU-FMI region, the induced Zeeman exchange field $h_0$, and the induced superconducting coherence length $\xi$. In particular we analyze the evolution of the subgap spectrum as a function of the length $L$. Interestingly, we have found that depending on the ratio $h_0/\Delta$, the emerging ABS can eventually cross below the Fermi energy at certain critical values $L_c$, and induce spin-and fermion parity-changing quantum phase transitions. We argue that this type of device 
constitute a promising highly-tunable platform to engineer subgap ABS.
\end{abstract}

\maketitle

\section{Introduction}
%have generated a surge of interest recently as they 
%can lead to intriguing and exotic quantum states at low temperatures . 
The interplay of superconductivity and magnetism at the microscopic scale has attracted a great deal of attention in recent years \cite{Balatsky_2006, Heinrich18_Review_single_adsorbates, Pawlak19_Review_on_MFs_in_magnetic_chains, Choi19_Review_atomic_spins_on_surfaces}. For instance, the Yu-Shiba-Rusinov (YSR) states   \cite{Yu65_YSR_states,Shiba68_YSR_states,Rusinov68_YSR_states} arising from the exchange interaction of an atomic magnetic moment in contact with a superconductor,
have been proposed as fundamental building blocks to engineer quantum devices with topologically non-trivial ground states. In particular, the so-called  “Shiba chains” (i.e., one-dimensional arrays of magnetic atoms deposited on top of a clean superconductor) are systems predicted to support Majorana zero-modes at the ends of the chain \cite{Nadj-Perdge13_Majorana_fermions_in_Shiba_chains,
Klinovaja13_TSC_and_Majorana_Fermions_in_RKKY_Systems, Li14_TSC_induced_by_FM_metal_chains}, and could be used in topologically-protected quantum computation schemes. Low-temperature scanning-tunneling microscopy (STM) experiments have confirmed the presence of intruiguing zero-energy end-modes   \cite{NadjPerge14_Observation_of_Majorana_fermions_in_Fe_chains, Ruby15_MBS_in_Shiba_chain,
Feldman16_High_resolution_Majorana_Shiba_chain,
Pawlak16_Probing_Majorana_wavefunctions_in_Fe_chains,
Ruby2017_Co_Chain_as_possible_Majorana_platform, Kim18_Tailoring_MBS_in_magnetic_atom_chains_on_SC, Jaeck2021_Review_Shiba_chains}. 
%In the STM spectra, isolated YSR states appear as sharp resonances in the differential conductance $dI/dV$, symmetrically located around  the Fermi level at energies within the superconducting gap, and localized spatially around the impurity  \cite{Yazdani97_YSR_states, Ji08_YSR_states, Iavarone10_Local_effects_of_magnetic_impurities_on_SCs, Ji10_YSR_states_for_the_chemical_identification_of_adatoms,Ruby15_Tunneling_into_localized_subgap_states_in_SC}.  Recent studies indicate that YSR states are the result of complex (and in many cases, antagonistic) interactions associated with non-trivial effects such as Kondo screening and quantum phase transitions \cite{Franke11_Competition_of_Kondo_and_SC_in_molecules, Bauer13_Kondo_screening_and_pairing_on_Mn_phtalocyanines_on_Pb, Hatter2017_Scaling_of_YSR_energies, Hatter15_Magnetic_anisotropy_in_Shiba_bound_states_across_a_quantum_phase_transition}, complex orbital and spatial structure of the YSR wavefunctions \cite{Menard2015, Ruby_2016, Choi2017_Mapping_the_orbital_structure_of_Shiba_states,Scherubl20_Large_spatial_extension_of_YSR_states}, multiple subgap states  attributed to higher angular momenta $\ell$ \cite{Rusinov68_YSR_states, Ji08_YSR_states} or single-ion anisotropies in the magnetic states of the adsorbate \cite{Hatter15_Magnetic_anisotropy_in_Shiba_bound_states_across_a_quantum_phase_transition, Zitko11_Effect_of_magnetic_anisotropy_on_Shiba_states}, etc.

Other systems where the competition of  superconductivity and magnetism at the nanoscale generates exotic subgap states are superconductor (SU)- ferromagnet (FM) heterostructures, such as SU-FM-SU Josephson junctions and SU-FM proximity devices \cite{Bergeret05_Review_SC_FM_heterostructures,Eschrig18_Review_ABS_in_SFS_junctions}. Subgap states generated in these structures are usually referred to as Andreev bound states (ABS). More recently, a novel class of hybrid device, i.e., semiconductor (SE) nanowire systems combined with superconductors and ferromagnetic insulator (FMI) materials have been fabricated using molecular-beam epitaxy  techniques \cite{Liu19_SM_FMI_SC_epitaxial_nanowires, Vaitiekenas21_ZBPs_in_FMI_SC_SM_hybrid_nanowires}. These SE-SU-FMI hybrid structures allow to build nanostructures with specific tailored properties which are impossible to obtain with the isolated individual components. 

\begin{figure}
    \centering
    \includegraphics[width=0.9\columnwidth]{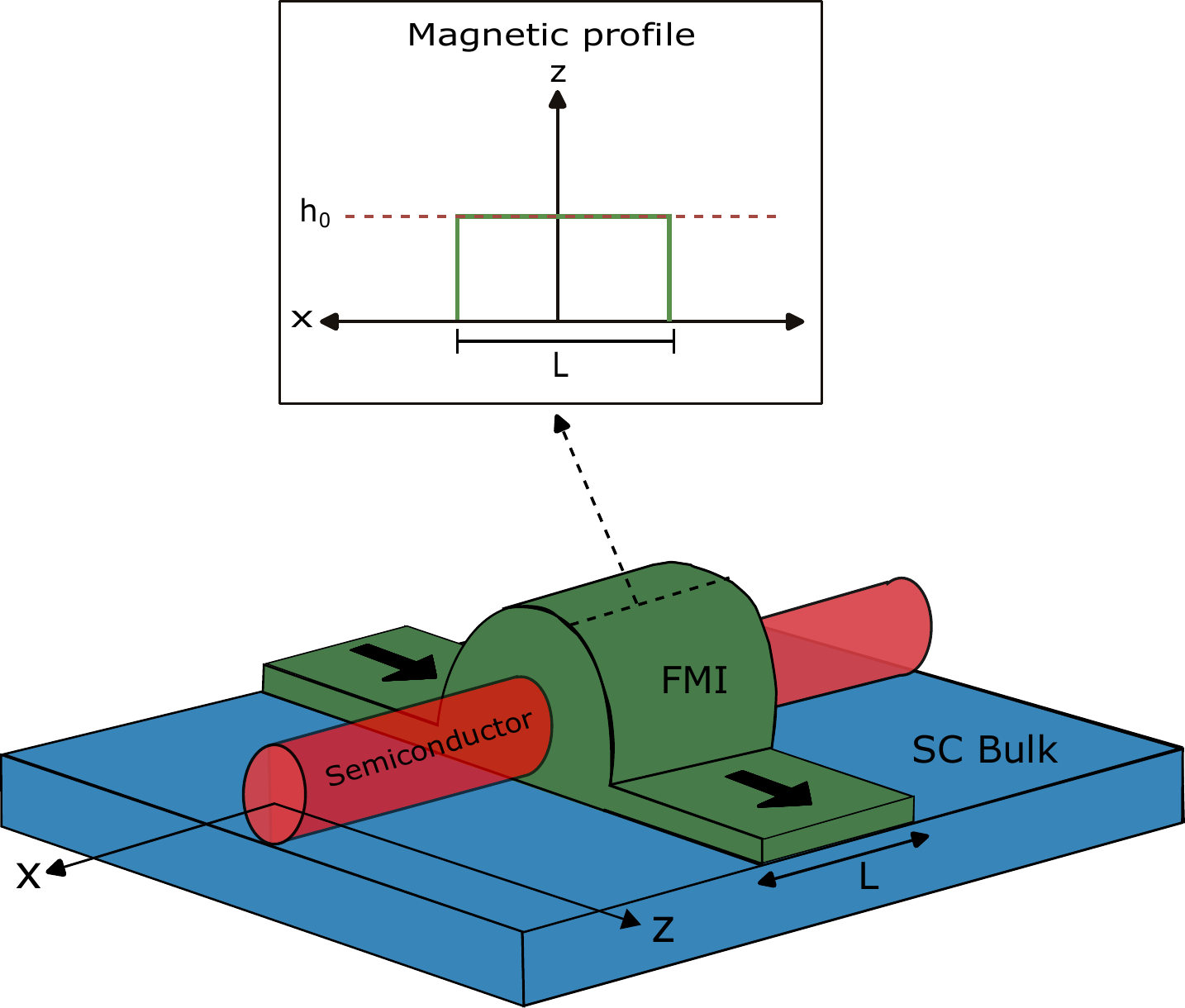}
    \caption{Schematic representation of the SC-FMI heterostructure.}
    \label{fig:system}
\end{figure}

Despite the evident differences between the abovementioned physical systems, from the theoretical perspective they can be described within  the same unified theoretical model combining superconductivity and local exchange fields at the microscopic scale. The emerging subgap states (which can be referred to as either YSR states or ABS, depending on the context) appear symmetrically around  the Fermi level $E_F$, and localize spatially around the impurity or the FM region. Their energy-position  within the gap depend on the value of the exchange field and on other experimental parameters. Interestingly, whenever one of these states crosses $E_F$, a spin- and parity-changing quantum phase transition, usually known as the “$0-\pi$” phase transition,  occurs \cite{Sakurai70, Balatsky_2006}. In the case of atomic ``Shiba impurities" or ultra-short SU-FM-SU junctions (i.e., junctions in which the length $L$ of the FM region is much smaller than $\lambda_F$, the Fermi wavelength of the superconductor \cite{Costa18_YSR_states_in_magnetic_JJs}), it is customary to consider the magnetic scatterer as a \emph{point-like} classical spin  $\mathbf{S}$ located at the point $\mathbf{R}_0$, interacting via a contact \textit{s-d} exchange interaction $H_\text{Z}=J(\mathbf{r})\;\mathbf{S}\cdot \mathbf{s}(\mathbf{r})$ with the host superconducting electrons \cite{Shiba68_YSR_states}. Here $J(\mathbf{r})=J_0 \delta(\mathbf{r}-\mathbf{R}_0)$ is the local exchange potential  and $\mathbf{s}(\mathbf{r})$ is the spin density vector of the electronic fluid. Subsequent theoretical works considered atomic-sized systems with finite- (albeit short-ranged) exchange interactions with spherical symmetry \cite{Rusinov68_YSR_states, Rusinov69_gapless_SC_due_to_magnetic_impurities, Flatte97_LDOS_of_defects_in_SCs, Arrachea21_Multiple_YSR_states_in_multiorbital_adatoms}. In that case, theory predicts the existence of multiple YSR states labelled by their orbital momentum $\ell$, a prediction  that has been recently observed in STM experiments \cite{Ji08_YSR_states, Ruby_2016,
Choi2017_Mapping_the_orbital_structure_of_Shiba_states}.

The behavior of subgap states and the associated $0-\pi$ quantum phase transitions has also been studied in the opposite limit $L\gg \lambda_F$ in the context of ballistic SU-FM-SU Josephson junctions with generic spin-dependent fields in the sandwiched region \cite{Konschelle16_ABS_in_SFS_junctions_PRL, Konschelle16_ABS_in_SFS_junctions_PRB,Rouco19_YSR_and_ABS_in_SFS_junctions}. In this case the results differ from the well-known results of YSR states due to the finite extension of the magnetic profile. In particular, the subgap spectrum of long SU-FM-SU junctions with zero phase difference is known to be double degenerate \cite{Eschrig18_Review_ABS_in_SFS_junctions, Konschelle16_ABS_in_SFS_junctions_PRB}, showing the inherent complexity of these hybrid heterostructures. On the experimental side, the possibility to engineer and control the position of the subgap states by a modification of the fabrication paramaters (e.g., the length $L$ or exchange field $h_0$ via different FM materials) opens interesting perspectives for potential electronic devices, where the precise knowledge of the subgap spectrum is crucial to control their transport properties.

Motivated by the experimental developments mentioned above, in this work we study the subgap states emerging in one-dimensional (1D) SE-SU-FMI heterostructures where the SU and the FMI layers simultaneously generate coexisting proximity-induced pairing and exchange interactions over a \emph{finite and arbitrary} length $L$ in the SE nanowire, as  schematically shown in Fig. \ref{fig:system}. This coexistence is a crucial aspect of this device, which makes it unique and different from the abovementioned SU-FM-SU junctions, where such overlap occurs only at the SU-FM interface. Our main goal in this work is to study and understand the behavior of the subgap ABS in this device as a function of the experimentally relevant parameters of the model, i.e., the length $L$ of the FMI region and the magnitude of the induced exchange field $h_0$. As mentioned above, a device similar to that shown in Fig. \ref{fig:system} has been recently experimentally realized in SE nanowires with epitaxially-grown SU and FMI layers \cite{ Liu19_SM_FMI_SC_epitaxial_nanowires, Vaitiekenas21_ZBPs_in_FMI_SC_SM_hybrid_nanowires}. While the main interest of that work was the fabrication of a device with non-trivial topological SU ground state hosting Majorana zero modes, here we will study the regime of parameters favoring a topologically-trivial ground state. As we will show below  (see Sec. \ref{sec:model}), this  case is already very complex and rich as a result of the antagonistic SU and FM interactions and, to the best of our knowledge, the detailed behavior of subgap states and the quantum phase diagram emerging in such a system have not been explicitly studied before. 

The article is organized as follows. In Section \ref{sec:model}, we introduce the model representing a 1D SE-SU-FMI hybrid nanowire, discuss the solution to the Bogoliubov-de Gennes equations for the subgap states, and derive a generic equation for the subgap spectrum. In Section \ref{sec:results}, we analyze the results in two specific limits, where we recover well-known results: a) the semiclassical limit, where the superconducting coherence length $\xi$ is much larger than the Fermi wavelength $\lambda_F$, and b) the atomic YSR limit, in which the exchange-field induced by the FMI region becomes a delta-function potential: i.e., infinitesimally narrow ($L\ll \lambda_F$), and infinitely deep ($h_0 \gg E_F$), in such a way that the product $h_0.L=J$ is kept constant. In both cases, well-known analytical solutions to the subgap spectrum can be recovered. In addition, we numerically solve the characteristic equation for the subgap states and provide a generic description of the subgap spectrum, not restricted to any of these limits. We find a rich behaviour of the subgap ABS, where the competing FM exchange and SU pairing interactions give rise to parity- and spin-changing quantum phase transitions. Finally, in Section \ref{sec:summary}, we present a summary and our conclusions.

\section{Theoretical model}\label{sec:model}
We focus on the system schematically depicted  in Fig. \ref{fig:system}, which represents a 1D SE-SU-FMI  hybrid  nanostructure of total length $L_\text{w}$, similar to those fabricated in Refs. \onlinecite{Liu19_SM_FMI_SC_epitaxial_nanowires, Vaitiekenas21_ZBPs_in_FMI_SC_SM_hybrid_nanowires}. We model this system with the Hamiltonian $H=H_{\text{w}}+H_{\Delta} +H_{\text{Z}}$, where
\begin{align}
H_{\text{w}} & =\sum_{\sigma}\int_{-\frac{L_{\text{w}}}{2}}^{\frac{L_{\text{w}}}{2}}dx\ \psi_{\sigma}^{\dagger}(x)\left[-\frac{\hbar^2\partial_{x}^2}{2m^*}-\mu\right]\psi_{\sigma}^{\dagger}(x),\label{eq:H_W}\\
H_{\text{ \ensuremath{\Delta}}} & =\Delta\int_{-\frac{L_{\text{w}}}{2}}^{\frac{L_{\text{w}}}{2}}dx\ \left[\psi_{\uparrow}^{\dagger}(x)\psi_{\downarrow}^{\dagger}(x)+\text{H.c.}\right],\label{eq:H_Delta}\\
H_{\text{Z}} & =\int_{-\frac{L_{\text{w}}}{2}}^{\frac{L_{\text{w}}}{2}}dx\ h(x)\left[\psi_{\uparrow}^{\dagger}(x)\psi_{\uparrow}(x)-\psi_{\downarrow}^{\dagger}(x)\psi_{\downarrow}(x)\right].\label{eq:H_Z}
\end{align}
Here $H_{\text{w}}$ is the Hamiltonian of a single-channel SE nanowire of length $L_\text{w}$, in which the fermionic operator $\psi_\sigma(x)$ creates an electron  at position $x$ with spin projection $\sigma=\uparrow , \downarrow$ and effective mass $m^*$.  The parameter $\mu$ is the chemical potential, which  can be experimentally varied applying external gates beneath the nanostructure. 

The terms $H_ \Delta$ and $H_{\text{Z}}$ represent, respectively,
the proximity-induced pairing interaction encoded by the parameter $\Delta$, and the Zeeman exchange interaction introduced by the FMI and described by a space-dependent  exchange field $h(x)$, which we assume oriented along the $z$ direction (see Fig. \ref{fig:system}). Moreover, since these interactions are externally induced into the semiconductor, we make the additional assumption that $\Delta$ is unaffected by the presence of $h(x)$ (a renormalized value of $\Delta$ does not change qualitatively our results). As mentioned before, these two terms can be effectively induced by the presence of epitaxially-grown SU and FMI shells in contact with the SE nanowire \cite{Liu19_SM_FMI_SC_epitaxial_nanowires, Vaitiekenas21_ZBPs_in_FMI_SC_SM_hybrid_nanowires}. It has been experimentally confirmed \cite{Vaitiekenas21_ZBPs_in_FMI_SC_SM_hybrid_nanowires} that the FMI shell (EuS in that case) consists of a single magnetic monodomain, and therefore modelling this layer by the Hamiltonian $H_{\text{Z}}$ is a reasonable approximation. In addition, the epitaxially-generated interfaces are essentially disorder-free, a necessary condition to produce a proximity-induced hard-gap \cite{Takei13_Soft_gap}. This feature allows to neglect the effects of disorder and considerably simplifies the theoretical description. The presence of both, a hard proximity-induced superconductor gap and an effectively induced Zeeman field, in these nanowires have been reported in transport measurements in Refs. \onlinecite{Liu19_SM_FMI_SC_epitaxial_nanowires, Vaitiekenas21_ZBPs_in_FMI_SC_SM_hybrid_nanowires}.  In addition, note that in the above model we have neglected the effect of the Rashba spin-orbit interaction. While this interaction is crucial for the emergence of a topologically non-trivial (i.e., D class) superconducting phase supporting Majorana zero-modes \cite{Sau15_Bound_states_in_a_FM_wire_in_SC}, here we will focus strictly on the topologically-trivial ground state. As we will show below, the competition of SU and FM interactions make this system already very complex and interesting in itself. 

We note that since the total single-particle fermionic spin along $z$
\begin{align}\label{eq:sz}
s_{z} & =\frac{1}{2}\int_{-\frac{L_{\text{w}}}{2}}^{\frac{L_{\text{w}}}{2}}dx\ \left[\psi_{\uparrow}^{\dagger}(x)\psi_{\uparrow}(x)-\psi_{\downarrow}^{\dagger}(x)\psi_{\downarrow}(x)\right],
\end{align}
is a conserved quantity which verifies $\left[s_z,H\right]=0$, we can label the electronic eigenstates of $H$ with $\sigma=\left\{\uparrow, \downarrow\right\}$. Therefore, we introduce the following Nambu spinors
\begin{align}\label{eq:nambu_spinors}
\Psi_{\uparrow}(x) & =\left(\begin{array}{c}
\psi_{\uparrow}(x)\\
\psi_{\downarrow}^{\dagger}(x)
\end{array}\right),\hspace{1em} & \Psi_{\downarrow}(x) & =\left(\begin{array}{c}
\psi_{\downarrow}(x)\\
\psi_{\uparrow}^{\dagger}(x)
\end{array}\right),
\end{align}
related to each other via the charge-conjugation transformation $\Psi_{\bar{\sigma}}(x)=\mathcal{K} \tau_x \Psi_{\sigma}(x)$, where $\tau_x$ is the $2\times 2$ Pauli matrix, and $\mathcal{K}$ is the complex conjugation operator. In terms of these spinors the Hamiltonian writes
\begin{align}\label{eq:H_BdG_operator}
H & =\frac{1}{2}\sum_{\sigma}\int_{-\frac{L_{\text{w}}}{2}}^{\frac{L_{\text{w}}}{2}}dx\ \Psi_{\sigma}^{\dagger}(x)\mathcal{H}_{\text{BdG,}\sigma}(x)\Psi_{\sigma}(x),
\end{align}
where the Bogoliubov-de Gennes (BdG) Hamiltonian is defined as
\begin{align}\label{BdG}
\mathcal{H}_{\text{BdG,}\sigma}& =\left(\begin{array}{cc}
-\frac{\hbar^2\partial_{x}^2}{2m}-\mu+\sigma h(x) & \sigma \Delta\\
\sigma \Delta & \frac{\hbar^2\partial_{x}^2}{2m}+\mu+\sigma h(x)
\end{array}\right).
\end{align}
In this expression, the spin projection $\sigma = \uparrow$ ($\downarrow$) on the left-hand side corresponds to the $+$ ($-$) sign in the definition of the BdG matrix. Using the above charge-conjugation transformation, we note that the BdG Hamiltonian Eq. (\ref{BdG}) verifies the following symmetry transformation 
\begin{align}\label{eq:BdG_symmetry}
\mathcal{K}\tau_x\mathcal{H}_{\text{BdG},\sigma}& =
-\mathcal{H}^*_{\text{BdG},\bar{\sigma}}\mathcal{K}\tau_x,
\end{align}
and therefore, provided $\chi_\sigma(x)$ is a solution of the BdG eigenvalue equation
\begin{align}\label{eq:BdG_eigenvalues}
\mathcal{H}_{\text{BdG},\sigma}(x)\chi_{\sigma}(x) & =E_{\sigma}\chi_{\sigma}(x),
\end{align}
with eigenenergy $E_\sigma$, the transformed spinor $\chi_{\bar\sigma}(x)=\mathcal{K} \tau_x \chi_{\sigma}(x)$, is also a solution
with eigenenergy $E_{\bar{\sigma}}=-E_{\sigma}$.

In what follows, we assume for simplicity the thermodynamic limit $L_\text{w}\rightarrow \infty$, and we focus on the features introduced by the magnitude and spatial dependence of $h\left(x\right)$, which is crucial for the rest of this work. In addition, we assume the following  step-like spatial profile for the exchange field
\begin{align}\label{eq:h_x}
h(x) & =\begin{cases}
-h_{0} & \text{if }\left|x\right|<\frac{L}{2},\\
0 & \text{if }\left|x\right|\geqslant \frac{L}{2},
\end{cases}
\end{align}
which models a uniform FMI shell of length $L$ in contact with the SE nanowire (see Fig. \ref{fig:system}). This choice for $h(x)$ allows to split the problem into regions with either $\left|x\right|<\frac{L}{2}$ or $\left|x\right|>\frac{L}{2}$,  with generic exponential solutions
\begin{align}\label{eq:ansatz}
\chi_{\sigma}(x) & \sim \left(\begin{array}{c}
\alpha_\sigma\\
\beta_\sigma
\end{array}\right)e^{ikx}.
\end{align}
Linear combinations of Eq. (\ref{eq:ansatz}), with appropriate coefficients and with allowed values of $k$ for each region, must be built so that continuity of the total wavefunction and its derivative  at the interfaces is satisfied. With this requirement, the solution of Eq.(\ref{eq:BdG_eigenvalues}) is finally obtained.

Note that the BdG Hamiltonian (\ref{BdG}) is even under space inversion $x\rightarrow -x$, and therefore its eigenstates must be even or odd under this transformation of coordinates. This symmetry allows to reduce the number of unknowns of the problem (i.e., coefficients of the linear combintation). Replacing the above ansatz Eq. (\ref{eq:ansatz}) into the BdG eigenvalue Eq. (\ref{eq:BdG_eigenvalues}), and looking for localized solutions with energy within the gap $\left| E_\sigma \right|<\Delta$, we obtain the following expressions for the eigenstates belonging to the even-symmetry subspace:
\begin{widetext}
\begin{align}
\chi_{e,\sigma}\left(x>\frac{L}{2}\right) & =A_{1\sigma}^{e}\left(\begin{array}{c}
1\\
\sigma e^{-i\varphi_{\sigma}}
\end{array}\right)e^{-\kappa_{\sigma}x}+A_{2\sigma}^{e}\left(\begin{array}{c}
1\\
\sigma e^{i\varphi_{\sigma}}
\end{array}\right)e^{-\kappa_{\sigma}^{\ast}x},\label{eq:chi_e_III}\\
\chi_{e,\sigma}\left(-\frac{L}{2}\leq x\leq\frac{L}{2}\right) & =B_{1\sigma}^{e}\left(\begin{array}{c}
1\\
\sigma e^{-\eta_{\sigma}}
\end{array}\right)\cos k_{\sigma}x+B_{2\sigma}^{e}\left(\begin{array}{c}
1\\
\sigma e^{\eta_{\sigma}}
\end{array}\right)\cos\bar{k}_{\sigma}x\label{eq:chi_e_II},
\end{align}
and the following expressions for the odd-symmetry eigenfunctions
\begin{align}
\chi_{o,\sigma}\left(x>\frac{L}{2}\right) & =A_{1\sigma}^{o}\left(\begin{array}{c}
1\\
\sigma e^{-i\varphi_{\sigma}}
\end{array}\right)e^{-\kappa_{\sigma}x}+A_{2\sigma}^{o}\left(\begin{array}{c}
1\\
\sigma e^{i\varphi_{\sigma}}
\end{array}\right)e^{-\kappa_{\sigma}^{\ast}x},\label{eq:chi_o_III}\\
\chi_{o,\sigma}\left(-\frac{L}{2}\leq x\leq\frac{L}{2}\right) & =B_{1\sigma}^{o}\left(\begin{array}{c}
1\\
\sigma e^{-\eta_{\sigma}}
\end{array}\right)\sin k_{\sigma}x+B_{2\sigma}^{o}\left(\begin{array}{c}
1\\
\sigma e^{\eta_{\sigma}}
\end{array}\right)\sin\bar{k}_{\sigma}x\label{eq:chi_o_II},
\end{align}
\end{widetext}
where the coefficients $\{A^\nu_{1\sigma},A^\nu_{2\sigma},B^\nu_{1\sigma},B^\nu_{2\sigma}\}$, with $\nu=\{e,o\}$, are unknowns to be fixed. In addition, in the above expressions we have introduced the parametrization
\begin{align}
\cos\varphi_\sigma & =\frac{E_\sigma}{\Delta}, \label{eq:parametrization_phi_sigma}\\
\cosh\eta_\sigma& =\frac{E_\sigma+\sigma h_0}{\Delta},\label{eq:parametrization_eta_sigma}
\end{align}
where we fix the definition of $\varphi_\sigma$ to the interval $\varphi_\sigma \in (0,\pi]$. The phase variable $\varphi_\sigma$ is associated to the Andreev reflection taking place at the interface $x_b=L/2$. Note that the parametrization in Eq. (\ref{eq:parametrization_eta_sigma}) makes sense whenever the right-hand side is positive. If this condition is not satisfied, one can always use the symmetry Eq.(\ref{eq:BdG_symmetry}) to send $E_\sigma \to -E_{\bar{\sigma}}$ and $\sigma \to \bar{\sigma}$. In addition, note that whenever $1\leq \left(E_\sigma+\sigma h_0\right)/\Delta$
the parameter $\eta_\sigma$ is purely real, while for  $0<\left(E_\sigma+\sigma h_0\right)/\Delta<1$ it is purely imaginary. 
Finally, we have introduced the quantities
\begin{align}
    \kappa_{\sigma} & \equiv-ik_{F}\sqrt{1+\frac{2i}{ k_{F}\xi}\sin\varphi_{\sigma},}\label{eq:kappa_sigma}\\
    k_{\sigma} & \equiv k_{F}\sqrt{1+\frac{2}{k_{F}\xi}\sinh\eta_{\sigma}},\label{eq:k_sigma}\\
    \bar{k}_{\sigma}&\equiv k_{F}\sqrt{1-\frac{2}{k_{F}\xi}\sinh\eta_{\sigma}},\label{eq:kbar_sigma}
\end{align}
and the definition of the coherence length of the (proximity-induced) 1D superconductor $\xi=\hbar v_F/\Delta$. Notice also that the spatial dependence of the wavefunctions in the region $x<-L/2$ can be readily obtained by symmetry from the relations $\chi_{e,\sigma}\left(x\right)=\chi_{e,\sigma}\left(-x\right)$, and $\chi_{o,\sigma}\left(x\right)=-\chi_{o,\sigma}\left(-x\right)$.

We can intuitively understand the form of the scattering solutions in the regions $x>L/2$ and $x<-L/2$ in the limit $k_F \xi \gg 1$ (i.e., the semiclassical limit, see Sec.\ref{sec:semiclassical}), where the momentum $\kappa_\sigma$ in Eq. (\ref{eq:kappa_sigma}) can be expanded as $\kappa_\sigma \simeq -i k_F + \sin \varphi_\sigma/k_F \xi$, and the  eigenfunctions Eqs. (\ref{eq:chi_e_III}) and (\ref{eq:chi_o_III}) take the form
\begin{align}\label{eq:chi_semiclassical}
    \chi_{\nu,\sigma}\left(x>\frac{L}{2}\right) & \approx \left[ A_{1\sigma}^{\nu}\left(\begin{array}{c}
1\\
\sigma e^{-i\varphi_{\sigma}}
\end{array}\right)e^{ik_F x}+\right.\nonumber\\
&\left.+A_{2\sigma}^{\nu}\left(\begin{array}{c}
1\\
\sigma e^{i\varphi_{\sigma}}
\end{array}\right)e^{-ik_F x}\right]e^{-\frac{\sin\varphi_\sigma x}{\xi}},
\end{align}
with $\nu=\{e,o\}$. In this way, it becomes evident that the component proportional to $A^\nu_{1\sigma}$ corresponds to a right-moving particle $\sim e^{i k_F x}$ while  $A^\nu_{2\sigma}$ corresponds to a left-moving particle $\sim e^{-i k_F x}$. In addition, the wavefunctions exponentially decay into the superconductor within a localization length $\lambda_\text{loc} = \xi/\sin\varphi_\sigma =\xi/\sqrt{1-(E_\sigma/\Delta)^2}$. These results are in complete agreement with Ref. \cite{Rouco19_YSR_and_ABS_in_SFS_junctions}, where the spectrum of SU-FM-SU Josephson junctions has been recently studied as a function of the length $L$ of the FM region. However, in our case, the presence of a finite pairing gap $\Delta$ in the region  $-L/2 < x< L/2$ (as opposed to the assumption $\Delta=0$ in the FM region in that work), gives rise to important differences which we analyze below in Sec. \ref{sec:results}. 

\subsection{Continuity conditions at the interface}

We now impose the continuity conditions on the wavefunction and its derivative at the boundary $x_b=L/2$:
\begin{align}
    \chi_{\nu,\sigma}\left(x_b^-\right)&=    \chi_{\nu,\sigma}\left(x_b^+\right)\label{eq:continuity_chi}\\
    \partial_x\chi_{\nu,\sigma}\left(x_b^-\right)&=    \partial_x\chi_{\nu,\sigma}\left(x_b^+\right)\label{eq:continuity_dchi}.
\end{align}
Note that the same equations are obtained by symmetry at the other boundary $-x_b$. Inserting the solutions Eqs. (\ref{eq:chi_e_III})-(\ref{eq:chi_o_II}), we can express the continuity equations in matrix form as
\begin{widetext}
\begin{align}
\begin{pmatrix}1 & \sigma e^{-i\varphi_{\sigma}}\\
\sigma e^{-i\varphi_{\sigma}} & 1
\end{pmatrix}\left(\begin{array}{c}
a_{1\sigma}^{\nu}\\
a_{2\sigma}^{\nu}
\end{array}\right) & =\begin{pmatrix}1 & \sigma e^{-\eta_{\sigma}}\\
\sigma e^{-\eta_{\sigma}} & 1
\end{pmatrix}\begin{pmatrix}F_\nu\left(\frac{k_{\sigma}L}{2}\right) & 0\\
0 & F_\nu\left(\frac{\bar{k}_{\sigma}L}{2}\right)
\end{pmatrix}\left(\begin{array}{c}
b_{1\sigma}^{\nu}\\
b_{2\sigma}^{\nu}
\end{array}\right),\label{eq:matrix_continuity_psi_even}\\
-\begin{pmatrix}1 & \sigma e^{-i\varphi_{\sigma}}\\
\sigma e^{-i\varphi_{\sigma}} & 1
\end{pmatrix}\begin{pmatrix}\kappa_{\sigma} & 0\\
0 & \kappa_{\sigma}^{\ast}
\end{pmatrix}\left(\begin{array}{c}
a_{1\sigma}^{\nu}\\
a_{2\sigma}^{\nu}
\end{array}\right) & =-s(\nu)\begin{pmatrix}1 & \sigma e^{-\eta_{\sigma}}\\
\sigma e^{-\eta_{\sigma}} & 1
\end{pmatrix}\begin{pmatrix}k_{\sigma}G_\nu\left(\frac{k_{\sigma}L}{2}\right) & 0\\
0 & \bar{k}_{\sigma}G_\nu\left(\frac{\bar{k}_{\sigma}L}{2}\right)
\end{pmatrix}\left(\begin{array}{c}
b_{1\sigma}^{\nu}\\
b_{2\sigma}^{\nu}
\end{array}\right),\label{eq:matrix_continuity_dpsi_even}
\end{align}
\end{widetext}
where we have conveniently redefined the unknown coefficients  as
\begin{align}
   A_{1\sigma}^{\nu} & \to e^{\kappa_{\sigma}L/2}a_{1\sigma}^{\nu} & B_{1\sigma}^{\nu} & \to b_{1\sigma}^{\nu}\\
A_{2\sigma}^{\nu} & \to\sigma e^{\kappa_{\sigma}^{\ast}L/2}e^{-i\varphi_{\sigma}}a_{2\sigma}^{\nu} & B_{2\sigma}^{\nu} & \to\sigma e^{-\eta_{\sigma}}b_{2\sigma}^{\nu},
\end{align}
in order to give these equations a more symmetric form.
In addition, we have used the notation $s(\nu)=+1(-1)$ for $\nu=e(o)$, and $F_e(x)=G_o(x)\equiv \cos(x)$, $G_e(x)=F_o(x)\equiv \sin(x)$ for compactness.
 
In each subspace (even or odd) we have four equations and four unknowns. Eliminating the variables $(b^\nu_{1\sigma},b^\nu_{2\sigma})^T$, and writing the equation for $(a^\nu_{1\sigma},a^\nu_{2\sigma})^T$, we find from the nullification of the corresponding determinant the following equations:
\begin{widetext}
\begin{align}
\frac{\cosh\eta_{\sigma}\cos\varphi_{\sigma}-1}{\sinh\eta_{\sigma}\sin\varphi_{\sigma}}&=
\begin{cases}
\dfrac{\vert\kappa_{\sigma}\vert^{2}-\left(K_{\sigma}+\bar{K}_{\sigma}\right)\Re\kappa_{\sigma}+K_{\sigma}\bar{K}_{\sigma}}{\left(\bar{K}_{\sigma}-K_{\sigma}\right)\Im\kappa_{\sigma}} & \text{(even-symmetry subspace)},\\
\\
\dfrac{\vert\kappa_{\sigma}\vert^{2}+\left(Q_{\sigma}+\bar{Q}_{\sigma}\right)\Re\kappa_{\sigma}+Q_{\sigma}\bar{Q}_{\sigma}}{\left(Q_{\sigma}-\bar{Q}_{\sigma}\right)\Im\kappa_{\sigma}} & \text{(odd-symmetry subspace)},
\end{cases}
\label{eq:main_equations}
\end{align}
\end{widetext}
where we have defined the quantities 
\begin{align}
    K_{\sigma}&=k_{\sigma}\tan\left(\frac{k_{\sigma}L}{2}\right),\label{eq:K_sigma}\\
    \bar{K}_{\sigma}&=\bar{k}_{\sigma}\tan\left(\frac{\bar{k}_{\sigma}L}{2}\right),\label{eq:bar_K_sigma}\\
    Q_{\sigma}&=k_{\sigma}\cot\left(\frac{k_{\sigma}L}{2}\right),\label{eq:Q_sigma}\\ 
    \bar{Q}_{\sigma}&=\bar{k}_{\sigma}\cot\left(\frac{\bar{k}_{\sigma}L}{2}\right).\label{eq:bar_Q_sigma}
\end{align}
From Eq. (\ref{eq:main_equations}), the eigenvalue $E_\sigma$ for each subspace is finally obtained. This equation summarizes our main theoretical results. In the next Sec. \ref{sec:results} we analyze the numerical solution and different important limits.

\subsection{Spin-changing quantum phase transitions}\label{sec:qpt}

We now focus on the quantum phase transitions which occur whenever one of the subgap states crosses $E_{F}$. To that end, let us analyze the spinors defined in Eq. (\ref{eq:nambu_spinors}), and consider the norm of the ``up" spinor
\begin{align*}
q_\uparrow  & =\int_{-L_{\text{w}}/2}^{L_{\text{w}}/2}dx\ \left[\psi_{\uparrow}^{\dagger}\left(x\right)\psi_{\uparrow}\left(x\right)+\psi_{\downarrow}\left(x\right)\psi_{\downarrow}^{\dagger}\left(x\right)\right].
\end{align*}
Recalling the definition of the single-particle $s_z$ operator [see Eq. (\ref{eq:sz})], it is straightforward to associate these two quantities through the relation $q_\uparrow=2 s_z -1$. Since $s_z$ is a conserved quantity, so is the norm $q_\uparrow$ of the ``up" Nambu spinors. This connection allows to interpret $q_\uparrow$ as an effective ``conserved charge". Similar considerations allow to write the relation $q_\downarrow=-2 s_z -1$. Due to the particle-hole relation Eq.(\ref{eq:BdG_symmetry}), the information about $s_z$ can be obtained with either $q_\uparrow$ or $q_\downarrow$. A more symmetric form involving both conserved charges is
\begin{align}\label{eq:sz_qup_qdown}
s_z&=\frac{q_\uparrow-q_\downarrow}{4}.
\end{align}
While redundant, this expression makes explicit that in the spin-symmetric case $q_\uparrow=q_\downarrow$, the net spin $s_z$ must vanish ($s_z=0$). 

We now return to Hamiltonian Eq. (\ref{BdG}),
and let us separate the effect of the proximity-induced Zeeman field, by writing it as $\mathcal{H}_{\text{BdG},\sigma}=\mathcal{H}_{0,\sigma}+\mathcal{V}_\sigma$, where
\begin{align}
\mathcal{H}_{0,\sigma} & =\begin{pmatrix}-\frac{\hbar^{2}\partial_{x}^{2}}{2m}-\mu & \sigma\Delta\\
\sigma\Delta & \frac{\hbar^{2}\partial_{x}^{2}}{2m}+\mu
\end{pmatrix},\label{eq:H0}\\
\mathcal{V}_\sigma & =\begin{pmatrix}\sigma h(x) & 0\\
0 & \sigma h(x)
\end{pmatrix}.\label{eq:V}
\end{align}
In this form, we can interpret the effect of the exchange field as a ``perturbation" on an otherwise homogeneous 1D superconductor represented by $\mathcal{H}_{0,\sigma}$.
Therefore, the full and the unperturbed single-particle Green's functions in this problem are respectively defined as
\begin{align}\label{eq:G_single_particle}
\mathcal{G}_\sigma\left(z\right)& =\left[z-\mathcal{H}_{0,\sigma} -\mathcal{V}_\sigma \right]^{-1},\\
\mathcal{G}_{0,\sigma}\left(z\right)& =\left[z-\mathcal{H}_{0,\sigma} \right]^{-1},
\end{align}
From here, the total number of effective ``up" charges $Q_\uparrow$ induced in the ground state due to the potential $\mathcal{V}_\sigma$, compared to the unperturbed homogeneous SU wire, can be computed as
\begin{align}
\Delta Q_\uparrow & =-\frac{1}{\pi}\text{Im}\ \text{Tr }\int_{-\infty}^{\infty}d\epsilon\ n_F\left(\epsilon\right)\Delta \mathcal{G}_\uparrow \left(\epsilon+i\delta\right).\label{eq:Delta_Q_eff}
\end{align}
where $\Delta \mathcal{G}_\sigma \left(z\right)\equiv  \mathcal{G}_\sigma\left(z\right)- \mathcal{G}_{0,\sigma} \left(z\right)$. At $T=0$, Eq. (\ref{eq:Delta_Q_eff})  can be easily computed from the well-known expression of the Friedel sum rule \cite{hewson}
\begin{align}
\Delta Q_\uparrow & =\frac{1}{\pi}\int_{-\infty}^{0}d\epsilon \left[\frac{\partial \eta_\uparrow\left(\epsilon\right )}{\partial \epsilon}-\frac{\partial \eta_{0,\uparrow}\left(\epsilon\right )}{\partial \epsilon}\right]  \\
&=  \frac{\eta_\uparrow\left(0\right ) - \eta_{0,\uparrow}\left(0\right )}{\pi}
\end{align}
where we have defined the phase shifts \cite{hewson,Rouco19_YSR_and_ABS_in_SFS_junctions}
\begin{align}
\eta_\sigma\left(\epsilon\right )&=\text{Im}\ln \det \mathcal{G}_\sigma\left(\epsilon+i\delta\right),\\
\eta_{0,\sigma}\left(\epsilon\right )&=\text{Im}\ln \det \mathcal{G}_{0,\sigma}\left(\epsilon+i\delta\right),
\end{align}
and where we have used that the phase shifts vanish in the limit $\epsilon \to \pm \infty$.

Since the system is non-interacting, the Green's function Eq. (\ref{eq:G_single_particle}) can be
written in terms of single-particle eigenstates $\left| \alpha,\sigma \right\rangle$, with $\alpha$ a generic label, as
\begin{align}\label{eq:G_eigenstates}
\mathcal{G}_\sigma\left(z\right) & =\sum_{\alpha}\frac{\left|\alpha,\sigma\right\rangle \left\langle \alpha,\sigma\right|}{z-E_{\alpha,\sigma}}. 
\end{align}
Therefore, after simple algebra, and using the above relations and  the fact that in the absence of magnetic field $s_z=0$ [see Eq. (\ref{eq:sz_qup_qdown})], the \textit{total} $S_z$ of the ground state is
\begin{align}\label{eq:Sz_quantized}
S_z&=\frac{\Delta Q_\uparrow}{2} =\frac{1}{2}\left[\sum_\alpha \Theta\left(-E_{\alpha,\uparrow} \right) - \sum_{\alpha^\prime} \Theta\left(-E^0_{\alpha^\prime, \uparrow} \right)\right],
\end{align}
where $\Theta(\epsilon)$ is the unit-step function. The above expression allows to interpret the total $S_z$ of the ground state as a function of the ``up" Nambu spinors with energy below $E_F=0$, as compared to the (unperturbed) situation $h_0=0$. Since the effective charges are quantized in integer numbers, the total spin $S_z$ can only change in discrete ``jumps" of 1/2 whenever a subgap state with projection up crosses below $E_F$ (note that we have defined dimensionless spin operators). This interpretation makes sense since the ground state becomes spin-polarized when the exchange field $h_0$ becomes large enough [i.e., the Zeeman energy of up-spin electron is \textit{decreased}, see Eqs. (\ref{eq:H_Z}) and (\ref{eq:h_x})]. While the result of Eq. (\ref{eq:Sz_quantized}) has been obtained recently by the authors of Ref. \cite{Rouco19_YSR_and_ABS_in_SFS_junctions}, we note that here we have rederived it in a different physical situation which allows a more generic regime of parameters. 

\section{Results}\label{sec:results}

We start this section by analyzing different limits of the general result given in Eq. (\ref{eq:main_equations}). In particular, in Sec. \ref{sec:semiclassical} we focus on the semiclassical limit, and in Sec. \ref{sec:ysr}  we study the atomic limit, where we recover the YSR results. In both cases, Eq. (\ref{eq:main_equations}) reduces to well-known analytical results. Finally in Sec. \ref{sec:generic} we show results corresponding to intermediate regimes, obtained by solving numerically Eq. (\ref{eq:main_equations}).

\subsection{Semiclassical limit}\label{sec:semiclassical}

Generally speaking, the semiclassical limit is verified when $E_F$ is the largest scale of the problem \cite{Duncan02_Semiclassical_theory_of_SCs}. In particular, the condition $E_F \gg \Delta $ (which is very well satisfied in most experimental systems) can be expressed as $ k_F \xi \gg 1$, recalling that after linearization of the normal quasiparticle dispersion, i.e., $\epsilon_{k,\sigma}\simeq \pm \hbar v_F k$, where the $+$($-$) sign corresponds to right-(left-)movers, the Fermi energy can be approximated as $E_F\simeq \hbar k_F v_F$.
In this case, Eqs. (\ref{eq:kappa_sigma})-(\ref{eq:kbar_sigma}) reduce to 
\begin{align}
    r_\sigma & \equiv \dfrac{\kappa_\sigma}{k_F}  \simeq -i  + \dfrac{\sin \varphi_\sigma}{k_F \xi},\\
    \zeta_\sigma &\equiv \dfrac{k_\sigma}{k_F}  \simeq 1 + \dfrac{\sinh \eta_\sigma}{k_F\xi},\\
    \bar{\zeta}_\sigma &\equiv \dfrac{\bar{k}_\sigma}{k_F} \simeq 1 - \dfrac{\sinh \eta_\sigma}{k_F\xi},
\end{align}
to leading order in $\mathcal{O}(k_F\xi)^{-1}$, and Eq. (\ref{eq:main_equations}) becomes 
\begin{align}
\frac{\cosh\eta_{\sigma}\cos\varphi_{\sigma}-1}{\sinh\eta_{\sigma}\sin\varphi_{\sigma}} &\simeq s(\nu)
\frac{1+\tan\left(\frac{k_{F}L\zeta_{\sigma}}{2}\right)\tan\left(\frac{k_{F}L\bar{\zeta}_{\sigma}}{2}\right)}{\tan\left(\frac{k_{F}L\zeta_{\sigma}}{2}\right)-\tan\left(\frac{k_{F}L\bar{\zeta}_{\sigma}}{2}\right)}, \nonumber \\
&=s(\nu) \cot\left({\dfrac{ L\sinh \eta_\sigma}{\xi}}\right),\label{eq:semiclassical_1}
\end{align}
where we have used the trigonometric identity $\tan{\left(x+y\right)}=(\tan\left(x\right) +\tan y)/(1+\tan{x}\tan{y})$. In general this transcendental equation cannot be solved analytically. However, in the regime of parameters $E_F\gg h_0\gg \Delta$, where the exchange field $h_0$ is much larger than $\Delta$, we can write $\cosh\eta_{\sigma}\approx\sinh\eta_{\sigma}\approx\left|\frac{h_{0}}{\Delta}\right|\gg 1$ [see  Eqs. (\ref{eq:parametrization_phi_sigma}) and (\ref{eq:parametrization_eta_sigma})
],
and Eq. (\ref{eq:semiclassical_1}) reduces to $\cot \varphi_\sigma =s(\nu) \cot\left( L h_0/\hbar v_F \right)$. Equivalently we can write this result as
\begin{align}
\arccos\left(\dfrac{E_{\sigma}}{\Delta}\right) & =\begin{cases}
\dfrac{LE_{\sigma}}{\hbar v_{F}}+\sigma\dfrac{Lh_{0}}{\hbar v_{F}}+2 n \pi,\quad\text{(even)}\\[5 pt]%\label{eq:rouco_even}\\
\\
\dfrac{LE_{\sigma}}{\hbar v_{F}}+\sigma\dfrac{Lh_{0}}{\hbar v_{F}}+\left(2n+1\right)\pi.\quad\text{(odd)}%\label{eq:rouco_odd}
\label{eq:semiclassical_2}
\end{cases}
\end{align}
This result can be interpreted as a semiclassical Bohr-Sommerfeld quantization condition for particles  which perform a complete a closed loop in the region $-L/2 <x<L/2$  \cite{Duncan02_Semiclassical_theory_of_SCs}. In particular, it
exactly coincides with theoretical results obtained for SU-FM-SU Josephson junctions with a normal (i.e., $\Delta=0$) FM region \cite{Rouco19_YSR_and_ABS_in_SFS_junctions, Konschelle16_ABS_in_SFS_junctions_PRB, Konschelle16_ABS_in_SFS_junctions_PRL}, the only difference being that within our theoretical treatment, we can distinguish the symmetry of the solutions. The similarity of these results can be rationalized noting that considering a normal sandwiched region in an SU-FM-SU junction corresponds to taking the limit $h_0 \gg \Delta$  in our Eq. (\ref{eq:semiclassical_1}) while keeping the ratio $E_\sigma/\Delta$ finite (since $E_\sigma$ corresponds to a subgap state, it is always bounded by $\Delta$), thus resulting in Eq. (\ref{eq:semiclassical_2}). This shows that our Eq. (\ref{eq:main_equations}) is a generic relation describing different situations regardless of the magnitude of the ratio $h_0/\Delta$.

\subsection{YSR-impurity limit}\label{sec:ysr}

We now consider the atomic YSR (or simply Shiba) limit, in which  the exchange profile becomes point-like, $L\to 0$, while $h_{0}\to\infty$, in such a way that the product $Lh_{0}=J=\text{const}$. Under these assumptions the magnetic barrier becomes a delta function and the Hamiltonian in Eq. (\ref{eq:H_Z}) can be written as
\begin{align}\label{eq:delta_function}
    H_{\text{Z}} & \approx -J \int_{-
    \infty}^{\infty}dx\ \delta(x)\left[\psi_{\uparrow}^{\dagger}(x)\psi_{\uparrow}(x)-\psi_{\downarrow}^{\dagger}(x)\psi_{\downarrow}(x)\right].
\end{align}
In this case, it is easy to see that the odd-symmetry solutions decouple from the above Hamiltonian (\ref{eq:delta_function}), as they vanish at $x=0$, and only even solutions can couple to the delta-potential.

As in the previous section, note that the limit $h_0 \to \infty$ implies $\cosh\eta_{\sigma}\approx\sinh\eta_{\sigma}\approx\left|\frac{h_{0}}{\Delta}\right|\gg 1$. However, the limit $h_0 \to \infty$ is not compatible with the semiclassical approach, as it violates the requirement $h_0\ll E_F$. Therefore we cannot use here our previous Eq. (\ref{eq:semiclassical_2}). Instead, we must first take  the limit $\eta_\sigma \gg 1$ together with the limit $L \to 0$, which applied to Eqs. (\ref{eq:k_sigma}) and (\ref{eq:kbar_sigma}) yield
 \begin{align}
 k_\sigma &\to k_F\sqrt{\dfrac{2h_0}{\hbar v_F k_F}},\\
 \bar{k}_\sigma &\to  i k_F\sqrt{\dfrac{2h_0}{\hbar v_F k_F}}.
 \end{align}
In addition Eqs. (\ref{eq:K_sigma})-(\ref{eq:bar_Q_sigma}) become
\begin{align}
    K_\sigma& \to\frac{k_{F}h_0 L}{\hbar v_{F}}=k_{F}\rho_{0}J,\\
    \bar{K}_\sigma& \to -\frac{k_{F}h_0 L}{\hbar v_{F}}=-k_{F}\rho_{0}J,
\end{align}
where the expressions for the density of states per spin of 1D quasiparticles at the Fermi energy $\rho_0=1/\hbar v_F$, and the exchange coupling $J=h_0 L$, have been used. Replacing these expressions into Eq. (\ref{eq:main_equations}) for the even-symmetry solutions, we obtain
\begin{align}
\sigma \dfrac{E^e_\sigma}{\sqrt{\Delta^2-\left(E^e\right)^2_\sigma}} & =\frac{1-\left(\rho_{0}J\right)^{2}}{\left(2J\rho_{0}\right)}.\label{eq:ysr_intermediate}
\end{align}
From this expression, we can easily solve for $E^e_\sigma$
\begin{align}
   \frac{E^e_{\sigma}}{\Delta}&=\sigma  \frac{1-\left(\rho_{0}J\right)^{2}}{1+\left(\rho_{0}J\right)^{2}},\label{eq:ysr_result}
\end{align}
which is the well-known expression for the energy of YSR-impurity subgap level \cite{Balatsky_2006}. This result indicates that any finite value of $J$ produces a YSR in-gap state. This type of subgap YSR states has been observed in several STM experiments on atomic magnetic adsorbates on superconducing substrates \cite{Yazdani97_YSR_states,Ji08_YSR_states, Iavarone10_Local_effects_of_magnetic_impurities_on_SCs, Ji10_YSR_states_for_the_chemical_identification_of_adatoms, Bauer13_Kondo_screening_and_pairing_on_Mn_phtalocyanines_on_Pb, Hatter15_Magnetic_anisotropy_in_Shiba_bound_states_across_a_quantum_phase_transition}. 

For completeness, and in order to illustrate the general scope of Eq. (\ref{eq:main_equations}), here we also show the result for the YSR odd states for a small (but finite) $L$. Using similar approximations, we obtain the expression
\begin{align}
   \frac{E^o_{\sigma}}{\Delta}&=\sigma  \dfrac{1}{\sqrt{1+\left(\dfrac{\rho_{0}J k_F^2L^2}{6} \right)^{2}}},\label{eq:ysr_odd_result}
\end{align}
where it becomes evident that in addition to a finite value of $J$, a finite value of $k_FL$ is needed to observe an odd-symmetry subgap YSR state.

\subsection{Subgap ABS spectrum in generic cases}\label{sec:generic}

As stated in  Section \ref{sec:model}, Eq. (\ref{eq:main_equations}) implicitly defines  the energy of the subgap states as a function of the parameters  $h_0/\Delta$ , $k_F \xi$, and $k_F L$. These parameters can be directly or indirectly controlled in experiments, i.e., the parameter $h_0$ can be controlled by modifying the FMI material, the length $L$ of the FMI region can be modified varying the length $L_\text{w}$ of the semiconductor via vapor-liquid-solid (VLS) method and subsequent evaporation of the FMI material \cite{Liu19_SM_FMI_SC_epitaxial_nanowires}, and the parameter $k_F$ in the semiconductor can be varied by changing the SE material or by introducing external gates to modify the chemical potential $\mu$. Therefore, due to this high degree of tunability, hybrid heterostructures might offer a unique platform to produce and control engineered subgap states. Probably the easiest way to experimentally control the subgap electronic structure is by producing different devices with the same FMI material and different lengths $L$. Therefore, in this section we show the numerical solutions of Eq. (\ref{eq:main_equations}) with fixed parameters $h_0/\Delta$ and  $k_F \xi$ (which control the ``operation regime" of the device), and calculate both the energy dependence of the even- and odd-symmetry ABS, and  the total spin $S_z$ of the device as a function of $L$ (i.e., dimensionless variable $k_F L$).

\begin{figure*}
\centering
\includegraphics{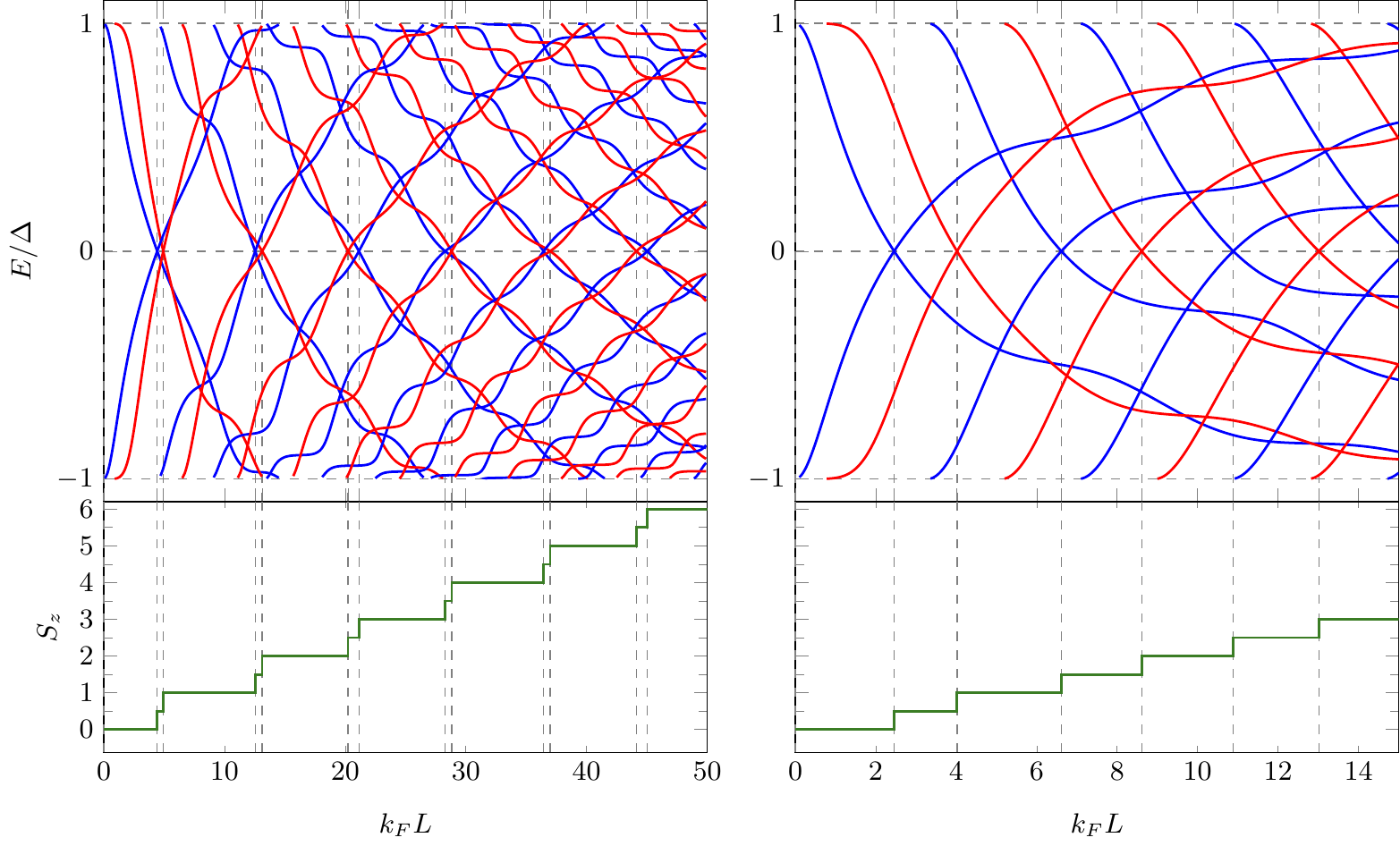}
\caption{Energy of the Andreev bound states (upper panel) and total spin $S_z$(lower panel) as a function of $k_F L$, for $k_F \xi = 7.8$ and $h_0/\Delta = 3.0$ (left panel) and $k_F \xi = 3.4$ and $h_0/\Delta = 2.1$ (right panel). Blue and red colors correspond to even and odd states respectively. Lines starting from the top gap edge at positive energy $E/\Delta=1$ (bottom gap edge at negative energy $E/\Delta=-1$) correspond to up (down) spin projections of the states. For smaller values of $k_F L$ (right panel), plateaus corresponding to regions of integer and half-integer spin are more separated and might become easier to observe in experiments.}
\label{Fig:Transitions}
\end{figure*}

Generally speaking, the overall evolution of the ABS spectrum from $L=0$ to $L\to \infty$ is quite complex and deserves a detailed explanation. As shown in Fig. \ref{Fig:Transitions}, as the parameter $k_F L$ increases, more and more subgap states emerge from the gap edges. This  behavior is  reminiscent of a quantum particle in a square-well potential, tipically taught in introductory quantum mechanics courses \cite{landau}, where increasing the width $L$ of the well increases the number of allowed bound states. In our case, the emergence of new ABS as $k_F L$ increases can be intuitively understood in terms of a competition between superconductivity and magnetic field: the magnetic field tends to break Cooper-pairs and to locally disrupt superconductivity in the magnetic region by introducing subgap states that become macroscopic in number for large $L$, eventually populating the whole gap. 

 We note that for any finite $L$,  even- and odd-symmetry states are generically non-degenerate (except at isolated points). However,  as it is clear from Figs. \ref{Fig:Transitions} and \ref{Fig:h0regions}, their energy difference (evidenced as oscillations of the blue and red lines around the semiclassical value) decreases very rapidly and the solutions become degenerate in the limit $L\to \infty$. This transition from non-degenerate YSR states in the limit $L\to 0$, to double degenerate ABS states for $L\to \infty$ has been discussed in previous works on ballistic SU-FM-SU junctions 
\cite{Eschrig18_Review_ABS_in_SFS_junctions, Rouco19_YSR_and_ABS_in_SFS_junctions, Konschelle16_ABS_in_SFS_junctions_PRB, Konschelle16_ABS_in_SFS_junctions_PRL}, and in the case of extended Shiba impurities in 1D nanowires \cite{Bortolin19_Bosonization_Shiba_impurity}. It is also clearly visible in Fig. \ref{Fig:Transitions}, and more dramatically in Fig.   \ref{Fig:h0regions} below. In our 1D geometry, this degeneracy in the limit $L\to \infty$ can be intuitively understood by linearizing the spectrum around the Fermi energy, and expressing the original fermionic operators in terms of right- and left-moving fields slowly varying in the scale of $k_F^{-1}$ \cite{giamarchi}, i.e., $\psi_\sigma\left(x\right) \approx e^{i k_F x} \psi_{R,\sigma}\left(x\right)+e^{-i k_F x} \psi_{L,\sigma}\left(x\right)$. The slowly-varying fields $\psi_{R,\sigma}(x)$ and $\psi_{L,\sigma}(x)$ are two independent chiral fermionic fields obeying the usual anticommutation relations, in terms of which the original Hamiltonian becomes \cite{Bortolin19_Bosonization_Shiba_impurity}
\begin{align}
H_{\text{w}} & \approx \sum_{\sigma}\int_{-\infty}^{\infty}dx\left[-i \hbar v_F \psi_{R,\sigma}^{\dagger}(x)\partial_x\psi_{R,\sigma}(x) \right. \nonumber \\ 
&+\left. i \hbar v_F\psi_{L,\sigma}^{\dagger}(x) \partial_x\psi_{L,\sigma}(x) \right]\\
H_{\text{ \ensuremath{\Delta}}} & \approx\Delta\int_{-\infty}^{\infty}dx\ \left[\psi_{R,\uparrow}^{\dagger}(x)\psi_{L,\downarrow}^{\dagger}(x)+\psi_{L,\uparrow}^{\dagger}(x)\psi_{R,\downarrow}^{\dagger}(x)+\text{H.c.}\right],\\
H_{\text{Z}} & \approx -\int_{-\infty}^{\infty}dx\ h_0\left[\psi_{R,\uparrow}^{\dagger}(x)\psi_{R,\uparrow}(x)-\psi_{R,\downarrow}^{\dagger}(x)\psi_{R,\downarrow}(x)\right.\nonumber \\
&+\left.\psi_{L,\uparrow}^{\dagger}(x)\psi_{L,\uparrow}(x)-\psi_{L,\downarrow}^{\dagger}(x)\psi_{L,\downarrow}(x)\right],
\end{align}
where oscillating terms proportional to $e^{\pm 2 i k_F x}$ have been neglected as they cancel out in the limit $L\to \infty$ due to destructive interference. Defining the new chiral Nambu spinors
\begin{align}\label{eq:nambu_spinors_linearized}
\Psi_{1,\sigma}(x) & =\left(\begin{array}{c}
\psi_{R,\sigma}(x)\\
\psi_{L,\bar{\sigma}}^{\dagger}(x)
\end{array}\right),\hspace{1em} & \Psi_{2,\sigma}(x) & =\left(\begin{array}{c}
\psi_{L,\sigma}(x)\\
\psi_{R,\bar{\sigma}}^{\dagger}(x)
\end{array}\right),
\end{align}
the Hamiltonian of the system can be expressed in terms of two decoupled chiral sectors
\begin{align}\label{eq:H_BdG_operator_linearized}
H & =\frac{1}{2}\sum_{\sigma=\uparrow,\downarrow}\sum_{j=1,2}\int_{-\infty}^{\infty}dx\ \Psi_{j,\sigma}^{\dagger}(x)\mathcal{H}_{j,\sigma}(x)\Psi_{j,\sigma}(x),
\end{align}
with the definitions of the chiral BdG Hamiltonians
\begin{align}\label{eq:H_BdG_matrix_linearized}
\mathcal{H}_{j,\sigma}& =\left(\begin{array}{cc}
(-1)^j i v_F \partial_x -\sigma h_0 & \sigma \Delta\\
\sigma \Delta & (-1)^{j+1} i v_F \partial_x -\sigma h_0 
\end{array}\right).
\end{align}
The Nambu spinors Eq. (\ref{eq:nambu_spinors_linearized}) define two independent chiral subspaces related by the inversion symmetry of the original Hamiltonian, i.e., under the space inversion operation $x\leftrightarrow -x$, the fermionic operators transform as $\psi_{L,\sigma}(x) \leftrightarrow \psi_{R,\sigma}(x)$, and consequently we conclude that $\Psi_{1,\sigma}(x) \leftrightarrow \Psi_{2,\sigma}(x)$, which must then be degenerate. In addition, the particle-hole symmetry Eq. (\ref{eq:BdG_symmetry}) in this representation produces $\Psi_{1,\sigma}(x) \rightarrow \Psi_{2,\bar{\sigma}}(x)$, and therefore $\mathcal{H}_{1,\sigma} \rightarrow -\mathcal{H}_{2,\bar{\sigma}}$, implying that the solutions verify the particle-hole symmetry property  $E_{1,\sigma}=-E_{2,\bar{\sigma}}$. Moreover, notice that assuming periodic boundary conditions, the problem can be solved with the solutions $\psi_{R,\sigma}(x)\sim e^{i k x}$ and $\psi_{L,\sigma}(x)\sim e^{-i k x}$, and the dispersion relation becomes $E_{1,\sigma}(k)=E_{2,\sigma}(k)=\pm \sqrt{(\hbar v_F k)^2+\Delta^2}-\sigma h_0$. From here, a renormalized quasiparticle gap  $2\Delta_\text{ren}=2\left|\Delta-h_0\right|$ is obtained, consistent with our previous result.

In terms of the chiral Nambu spinors, the most general solution is the linear combination 
\begin{align}
    \Psi_{\sigma}(x) & = Ae^{i k_F x}\Psi_{1,\sigma}(x) +Be^{-i k_F x}\Psi_{2,\sigma}(x).
\end{align}
This is exactly the same form that can be obtained by combining the degenerate even and odd solutions in Eqs.  (\ref{eq:chi_e_II}) and (\ref{eq:chi_o_II}) in the semiclassical limit where $k_F \xi \gg 1$.

From the analysis of the linearized Hamiltonian, we conclude that the  degeneracy in the limit $L\to\infty$ arises from the absence of chirality-breaking terms, i.e., terms $\sim \Psi_{1,\sigma}^\dagger(x)\Psi_{2,\sigma}(x)$ arising from, e.g., single particle backscattering terms $\psi^\dagger_{R,\sigma}(x)\psi_{L,\sigma}(x)$ or Cooper-pairing channels $\psi^\dagger_{R(L),\uparrow}(x)\psi^\dagger_{R(L),\downarrow}(x)$ carrying momentum $\mp 2k_F$. For this to occur, the magnetic FMI region must be uniform and its length $L$ must be much larger than $k_F^{-1}$ in order to produce the required cancellation of the  rapidly oscillating exponentials $\sim e^{\pm 2 i k_F x}$. In other words, the product $k_F L$ must be $k_F L\gg 1$, consistent with our numerical results in Figs. \ref{Fig:Transitions} and \ref{Fig:h0regions}. Only for small values of $k_F L$, where this destructive interference is incomplete, residual couplings of the type $\sim \Psi_{1,\sigma}^\dagger(x)\Psi_{2,\sigma}(x)$ remain, and the degeneracy is lifted.
Finally, we stress that the degeneracy in the limit $L\to \infty$ is a robust property to the presence of interactions, as shown in previous works \cite{Bortolin19_Bosonization_Shiba_impurity}.

On the other hand, in the limit $L\to 0$ and for any finite value of the Zeeman field $h_0$, both (even and odd) solutions converge to $E_\sigma/\Delta \to \pm 1$, indicating that the FMI region is no longer relevant (i.e., it physically drops from the description). However, the behavior near $L=0$ is quite different for each case: while the even-symmetry solution tends  to $E/\Delta \to 1$ as [see Eq. (\ref{eq:ysr_result})]
\begin{align}
    \frac{E^e_\sigma}{\Delta}&\approx \sigma\left[1 -2 \left(\frac{h_0 L}{\hbar v_F}\right)^2 \dots\right],
\end{align} 
from Eq. (\ref{eq:ysr_odd_result}) we conclude that the odd solution behaves as
\begin{align}
    \frac{E^o_\sigma}{\Delta}&\approx \sigma \left[1 -\frac{1}{2} \left(\frac{h_0 k_F^2 L^3}{6 \hbar v_F}\right)^2 \dots\right],
\end{align}
therefore approaching the gap edge much faster as $L\to 0$. 

%We now discuss the evolution of the subgap spectrum the hybrid device  as $L$ increases for the different regimes of $h_0/\Delta$. 

Besides the general features of the spectrum discussed up to this point, its evolution as $L$ increases is strongly affected by the values of the parameters $k_F\xi$ and $h_0/\Delta$. In what follows, we analyze their effects on Figs. \ref{Fig:Transitions} and Fig. \ref{Fig:h0regions} respectively.

\subsubsection{Effect of varying the parameter $k_{F}\xi$}

This parameter can be considered as  a ``knob'' which tunes the device from the semiclassical behavior ($k_F\xi$ large, see left panel in Fig. \ref{Fig:Transitions}) into a ``quantum'' regime ($k_F\xi$ small, see right panel) where the spectrum is dominated by quantum oscillations. The hybrid heterostructure under study is promising in this sense since, due to the combination of materials (in particular, semiconductors with a much smaller $k_F$ as compared to metals), it is in principle possible that $k_F\xi$ can be experimentally controlled. In addition, $k_F$ could be further modified by introducing external gating leads (through the modification of the chemical potential $\mu$).  To illustrate the dramatic changes in the spectrum as $k_F\xi$ varies, in Fig \ref{Fig:Transitions} we show the numerically obtained subgap spectra as a function of $k_FL$ for $k_F \xi = 7.8$ and $h_0/\Delta = 3.0$ (left panel), for and $k_F \xi = 3.4$ and $h_0/\Delta = 2.1$ (right panel). Solid blue (red) lines correspond to even(odd)-symmetry solutions. Moreover, since we always assume $h_0>0$, solutions emerging from the top edge $E/\Delta=1$ (bottom edge $E/\Delta=-1$) correspond to spin up (spin down) solutions. In addition, note the reflection symmetry of the solutions around the horizontal $E=0$ axis, a consequence of the particle-hole symmetry of the BdG Hamiltonian, Eq. (\ref{eq:BdG_symmetry}).

Upon decreasing $k_F\xi$, the subgap spectrum becomes much more intricate due to the enhanced even-odd energy-splitting, which results in an amplified oscillatory behavior of the ABS (we have reduced the range of $k_FL$ in the right panel for clarity in the figure). Unfortunately, in the regime  $k_F \xi \sim 1$ no analytic expressions for the subgap ABS are possible, but qualitative considerations can be provided. In fact, the amplified oscillations can be traced back to the larger energy dependence of the momenta Eq. (\ref{eq:kappa_sigma})-(\ref{eq:kbar_sigma}) as $k_F\xi$ decreases. Then, whereas for large $k_F\xi$ all these quantities converge to a static (i.e., energy-independent) value $\sim k_F$, the limit of small $k_F\xi$ produces a larger effect on the space-dependence of the wave functions through the exponential factors in Eqs. (\ref{eq:chi_e_III})-(\ref{eq:chi_o_II}). This in turn produces larger interference effects, and an enhanced lifting of the even-odd degeneracy.

This phenomenological behavior enables interesting possibilities, such as the chance to observe half-integer spin (and fermion parity-switching) quantum phase transitions in the ground state. To illustrate this effect, we show the ground-state $S_z$ transitions in the bottom panels of Fig. \ref{Fig:Transitions} in each case. While for larger $k_F\xi$, the half-integer $S_z$ steps are very narrow due to the almost-degenerate even-odd solutions (i.e., the even and odd solutions cross zero energy almost at the same value of $k_FL$), for smaller $k_F \xi$ the $S_z$ transitions occur in well-defined half-integer steps. This behavior is well explained by the enhanced lifting of the even-odd degeneracy, which allows to observe one ABS crossing zero energy at a time.% When $k_F \xi$ increases compared to $h_0/\Delta$ the even and odd states get closer to each other, making the observation of these parity-changing transitions more difficult. In contrast when the value of $k_F \xi$ is closer to $h_0/\Delta$ the even and odd states crossings separate, giving rise to half-integer regions of $S_z$ with similar size than the integer regions  Fig[\ref{Fig:Transitions}].

%For $h_0 > \Delta$ the states are allowed to cross zero energy. zero energy crossings that change the total spin in the system in $\frac{1}{2}$. In Fig[\ref{Fig:Transitions}] we have the evolution of the subgap states for $k_F \xi > h_0$,  we observe that the the distance between the zero crossing of the even and odd states is small, so $S_z$ have short regions of $k_F L$ with semi integer values therefore small region with change of parity. In comparison when the value of $k_F \xi$ is closer to $h_0$ the even and odd states separates to each other, resulting in semi integer regions of $S_z$ with similar length to the integer regions  Fig[\ref{Fig:Transitions}].

%%%%

\subsubsection{Effect of varying the parameter $h_0/\Delta$}

\begin{figure}
\centering
\includegraphics{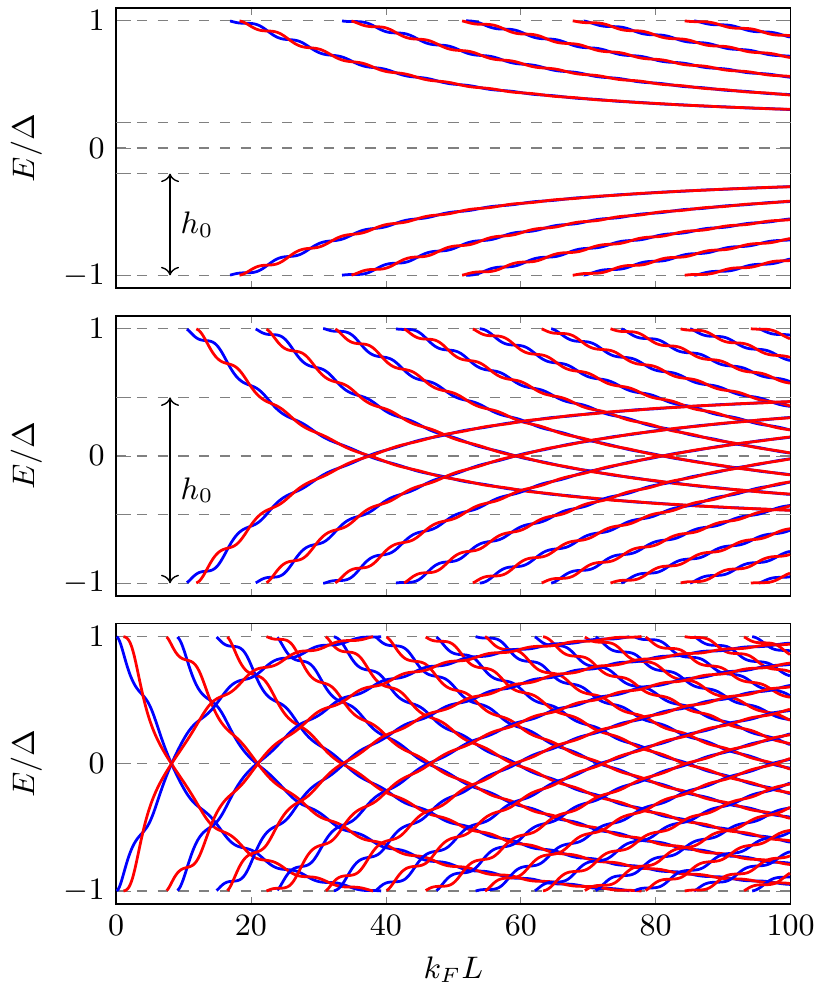}
\caption{Energy of the Andreev bound states as a function of $k_FL$ for the three different values of $h_0$ ($h_0/\Delta=0.8,\,1.54,\,2.2$ for the lower, middle and upper panels) and $k_F\xi = 8.2$. Blue and red colors correspond to even and odd states respectively. Lines starting from negative (positive) energies correspond to down (up) spin projections of the state. Note that the value of $h_0/\Delta$ sets the asymptotic limit for the Andreev states and is crucial to determine the overall subgap spectrum.}
\label{Fig:h0regions}
\end{figure}

In Fig. \ref{Fig:h0regions} we show the evolution of the subgap spectrum  as a function of $k_F L$,  for different values of the Zeeman field $h_0/\Delta=0.8, 1.54$ and $2.2$, and for a fixed relatively large value $k_F \xi=8.2$, allowing to interpret these results in terms of the semiclassical approximation.  Here we can clearly distinguish three qualitatively different regimes:  a) the ``weak field" regime $h_0 < \Delta$ (top panel) where the ABS  do not cross $E=0$, b) the ``intermediate field" regime $\Delta<h_0 < 2\Delta$ (middle panel) where the ABS can evenually cross zero energy, and quantum phase transitions can be induced, and finally c) the ``strong field" ($2\Delta<h_0$) regime (bottom panel), where the ABS can be found anywhere in the region $-1<E_\sigma/\Delta <1$. In all cases, the value of $h_0$ determines the asymptotic limit to which the ABS approach for large $L$ (see dashed black lines in Fig. \ref{Fig:h0regions}). Below we briefly discuss the main features of the spectrum in each regime.

\vspace{0.5cm}
\textit{a. Weak-field regime $0<h_0<\Delta$:} This regime is characterized by a Zeeman field which is not strong enough to destroy the superconducting gap. In this case none of the ABS is able to cross $E=0$ and in the limit $L\to \infty$ they asymptotically approach the value
$E_{\sigma}/\Delta \to \sigma \left(1 - h_0/\Delta\right)$ (see horizontal dashed black lines), and therefore a renormalized gap remains (see top panel in Fig. \ref{Fig:h0regions}).  More quantitatively, in the semiclassical limit [Eq. (\ref{eq:semiclassical_1})] they obey the asymptotic expression valid for $k_FL\rightarrow \infty$ 
\begin{align}
\frac{E^\nu_{\sigma}}{\Delta} & \simeq\sigma\left[1-\frac{h_{0}}{\Delta}+\frac{\pi^2}{2}\left(\frac{\xi}{L}\right)^{2}\left(1-\frac{s(\nu)\xi}{L}\sqrt{\frac{2\Delta}{h_0}-1}\right)^{2}\right],\label{eq:asymptotic_limit}
\end{align}
with $s(\nu)=1(-1)$ for $\nu=e(o)$. From here, we can clearly see that whereas the even-odd averaged quantities (i.e., the semiclassical values) approach the asymptotic limit as $L^{-2}$,  the energy difference between even and odd solutions (i.e., the amplitude of the oscillation around the semiclassical limit) decreases as $L^{-3}$, and the solutions become degenerate in the limit $L\to \infty$.
On the other hand, the quasiparticle gap in the limit $L\to \infty$ is renormalized to $2\Delta_\text{ren} = 2\left|\Delta - h_0\right|$. Note that this gap renormalization is quite specific to this setup, and is not  present, for instance, in the case of Ref. \cite{Rouco19_YSR_and_ABS_in_SFS_junctions}, where the magnetic region is normal and not superconducting, and in addition the system corresponds to a ``short" SU-FM-SU junction with $L<\xi$, and therefore only few subgap states are allowed.

Another feature of the weak-field regime is that the ABS require a minimal length $L_\text{min}$ to emerge in the subgap region. This can be easily understood in terms of Eq. \ref{eq:semiclassical_2}, where a minimal magnetic phase, represented by the product $Lh_0/\hbar v_F$, must be accumulated in order to produce an observable in-gap ABS. Finally, concerning the spin quantum number of the ground state, since none of the ABS cross $E_F$, no quantum phase transitions are expected according to the results of Sec. \ref{sec:qpt} and the value of the ground state spin remains a spin-singlet $S_z=0$.

\vspace{0.5cm}
\textit{b. Intermediate field regime $\Delta<h_0<2\Delta$:} In this case the Zeeman field $h_0$ is sufficiently strong to force the ABS to cross zero energy, eventually inducing quantum phase transitions (see middle panel in Fig. \ref{Fig:h0regions}). 
The $n$-th critical value $L_{\text{c},n}$ can be obtained imposing the condition $E_\sigma=0$ on the semiclassical approximation in Eq. (\ref{eq:semiclassical_1}), 
\begin{align}
L^\nu_{\text{c},n} & =\xi\frac{\arctan\left(-s\left(\nu\right)\mp\sqrt{\left(\frac{h_{0}}{\Delta}\right)^{2}-1}\right)+n\pi}{\sqrt{\left(\frac{h_{0}}{\Delta}\right)^{2}-1}}\label{eq:Lcritical},
\end{align}
with $s(\nu)=1(-1)$ for $\nu=e(o)$. 

In this regime, the ABS follow the same asymptotic behavior as in Eq. (\ref{eq:asymptotic_limit}), approaching $E_{\sigma}/\Delta \to \sigma \left(1 - h_0/\Delta\right)$,
although the overall subgap spectrum is completely different due to the closing of the gap, and due to the overlap of the $E_\uparrow$ and $E_\downarrow$ spectrum as $L$ increases beyond the first critical $L_{\text{c},0}$. In fact, in the regime $L>L_{\text{c},0}$ the quasiparticle gap becomes completely populated (and washed away) by subgap states. Moreover, we predict an accumulation of levels in the region $-\Delta+h_0<E<\Delta -h_0$, which can eventually form a peak structure in the total density of states.

\vspace{0.5cm}
\textit{c. Strong field regime $2\Delta<h_0$:} Finally, in this regime (see bottom panel in Fig. \ref{Fig:h0regions}), the asymptotic dashed lines fall within the continuum and it is no longer possible to obtain 
an analytic expression for the ABS behavior in the limit $L\to \infty$. As a result, the subgap ABS can be found anywhere in the subgap region $-1<E_\sigma/\Delta<1$. In addition, we note that the minimal length required to observe in-gap ABS has reduced to $L_\text{min}\approx 0$.

%\begin{widetext}

%\end{widetext}

\section{Summary and conclusions}\label{sec:summary}

In this work we have analyzed the subgap electronic structure in the one dimensional SE-SU-FMI  heterostructure schematically depicted in Fig. \ref{fig:system}, a novel physical system  recently fabricated using molecular beam epitaxy techniques (MBE). The main motivation to study this type of hybrid systems is that, via a careful combination of different materials, the emergent characteristics can be completely different from those of the individual components, providing a way to build devices with tailored properties and specific functionalities. In particular, much of the experimental effort has focused on the realization of topological superconducting phases hosting Majorana zero modes, with possible applications in topological quantum computing \cite{Liu19_SM_FMI_SC_epitaxial_nanowires, Vaitiekenas21_ZBPs_in_FMI_SC_SM_hybrid_nanowires}. A distinguishing feature of these heterostructures is the coexistence of antagonistic superconductor and ferromagnetic insulating layers over a finite and arbitrary length $L$ in a semiconductor wire, a combination that  confers unique spectral properties which cannot be found in elemental materials in nature. 

In particular, we have modelled the hybrid structure assuming non-interacting fermions in a one-dimensional single-channel nanowire under the effect of two proximity-induced interactions: a SU pairing and a space-dependent Zeeman exchange coupling [see Eqs. (\ref{eq:H_W})-(\ref{eq:H_Z})]. We have solved the associated Bogoliubov-de Gennes equations and, by imposing standard continuity conditions on the wave functions,  we have obtained an equation [Eq. (\ref{eq:main_equations})] defining the subgap ABS spectrum of the device. This single equation encodes our main theoretical results. We stress that our approach is equivalent to other works using the scattering-matrix formalism.  We have analytically solved Eq. (\ref{eq:main_equations})  in two paradigmatic limits: the semiclassical limit (Sec. \ref{sec:semiclassical}) and the Yu-Shiba-Rusinov limit, typical of atomic magnetic moments interacting with a superconductor (Sec. \ref{sec:ysr}). In both cases, we have been able to recover well-known analytical results, providing important sanity checks for our theoretical results. As a consequence of the symmetries of the Hamiltonian (i.e., inversion $x\to -x$ and $s_z$ spin  symmetries), it was possible to classify the solutions into even- and odd-symmetry, and with $s_z$ labels $\sigma=\uparrow, \downarrow$. In particular, we note that the even-odd classification, arising in the present case due to the inversion symmetry of the Hamiltonian, is nothing but the 1D analog of the classification in angular momentum eigenstates $\ell$ occurring in 3D spherically-symmetric Hamiltonians \cite{Rusinov68_YSR_states, Flatte97_LDOS_of_defects_in_SCs, Arrachea21_Multiple_YSR_states_in_multiorbital_adatoms}.

We have studied the subgap spectrum of ABS as a function of different parameters, namely: the length of the magnetic region (through the dimensionless parameter $k_FL$), the strength of the Zeeman exchange induced by the FMI (parameter $h_0/\Delta$), and the superconducting coherence length (parameter $k_F\xi$). We stress that each one of these parameters could in principle  (directly or indirectly) be controlled in experiments. However, due to its potential relevance for on-going experimental efforts, we have in particular focused our study on the evolution of the subgap spectrum as a function of the length $L$ (i.e., as it is probably the easiest parameter to vary in experiments), for fixed parameters $k_F\xi$ and $h_0/\Delta$. The parameter $L$ can be controlled by, e.g., changing the experimental  growing conditions of the semiconductor nanowires using the VLS growth method. In Figs. \ref{Fig:Transitions} and \ref{Fig:h0regions}  we have analyzed the evolution of the subgap spectrum in terms of the parameter $k_FL$ for different values of $h_0/\Delta$ and $k_F\xi$. Roughly speaking, while $k_F \xi$ controls the ``semiclassical vs quantum" operation regime of the device, and the magnitude of the even-odd energy separation, the parameter $h_0/\Delta$ essentially controls the energy separation of the $E_\uparrow$ and $E_\downarrow$ solutions, eventually enabling many interesting physical phenomena such as the possibility to observe multiple ABS crossing zero-energy, the existence of multiple spin- and parity-changing quantum phase transitions in the device, 
 quasiparticle gap renormalization $\Delta \to \Delta_\text{ren}=\left|\Delta-h_0\right|$ in the limit of large $k_F L$, etc.. An important conclusion here is that in order to experimentally observe a quantum phase transition, the condition $h_0 > \Delta$ must be fulfilled.  

Interpreting $L$ as a ``tunable" parameter has another theoretical advantage, as it enables to address the interesting fundamental question of how to connect two paradigmatic limits in SU-FM hybrid devices: the atomic limit ($k_FL\to 0$), where the physics is that of the well-known non-degenerate YSR states, and the ballistic limit ($k_FL\gg 1$) where the spectrum of the subgap ABS becomes double degenerate. Until very recently, these limits were treated as  disconnected from each other. In Ref. \cite{Rouco19_YSR_and_ABS_in_SFS_junctions} this issue  was addressed in the particular case of SU-FM-SU junctions in the limit $L< \xi$. Here  we have revisited this intriguing question for a different setup where such constraint does not exist, and have studied the evolution of the subgap spectrum as a function of $L$. The abovementioned symmetry classification into even and odd solutions is critically important to allow the interpretation of the degeneracy in the limit $k_FL \to \infty$ as an ``even-odd degeneracy". At the same time, it enables to explain the degeneracy lifting in the limit $L \to 0$, where only even states prevail in the subgap region of energies. Using an approximate model of one-dimensional fermions with linearized dispersion, we have provided a simple picture where the even-odd degeneracy naturally emerges as a consequence of destructive interferences of terms $e^{\pm i 2 k_F x}$ arising from single-particle backscattering mechanisms.    

The continuous evolution of the subgap spectrum as a function of $k_FL$ allows a better understanding of previous experimental STM results on atomic magnetic adsorbates on superconducting substrates, where the subgap YSR states are usually interpreted in terms of a point-like magnetic moment  \cite{Yazdani97_YSR_states,Ji08_YSR_states, Iavarone10_Local_effects_of_magnetic_impurities_on_SCs, Ji10_YSR_states_for_the_chemical_identification_of_adatoms, Bauer13_Kondo_screening_and_pairing_on_Mn_phtalocyanines_on_Pb, Hatter15_Magnetic_anisotropy_in_Shiba_bound_states_across_a_quantum_phase_transition}. While the delta-function limit is obviously  a mathematical idealization, in terms of our model the observed YSR states can be rationalized assuming a finite value of $k_F L$  and a (more physically appealing) finite value of the atomic local field $h_0$. This is precisely the case if we note that for magnetic impurities (e.g. Fe, Co or Mn atoms) deposited on top of bulk metallic S surfaces (e.g., Pb or Al), the spatial extension of the short-ranged Zeeman field can be estimated as the size of the $d$-shell orbitals $L\sim$ 1 \AA, while the Fermi wavevector of bulk superconductors (e.g., Pb) is $k_F \sim 1-2 \times 10^{10} \text{m}^{-1}$ (see Ref. \cite{ashcroft}). This type of adsorbate/substrate combination yields a parameter $k_FL\sim 1$, which is within the regime where we recover observable subgap states (see Figs. \ref{Fig:Transitions} and \ref{Fig:h0regions}).
On the other hand, in 1D semiconductor heterostructures as those of Refs. \onlinecite{Liu19_SM_FMI_SC_epitaxial_nanowires, Vaitiekenas21_ZBPs_in_FMI_SC_SM_hybrid_nanowires}, $k_F$ is usually much smaller than in metallic superconductors. Measurements of the number of carriers from the Hall conductance $R_H$ in 2D InGaAl quantum wells \cite{Kjaergaard15_PhD_Thesis_Proximity_induced_SC_in_low_dimensional_semiconductors} yield the estimated value $k_F \sim 2.2 \times 10^7 \text{m}^{-1}$, three orders of magnitude smaller as compared to bulk Pb. This much smaller value of $k_F$ allows for much larger, experimentally accessible values of $L$, while keeping values of $h_0$ also within experimental reach. All together, this combination makes these hybrid materials a much more versatile platform to control the spectrum of YSR/ABS subgap states.

To characterize the quantum phase transitions occurring in the device, we have computed the value of the total $S_z$ using a spin version of the Friedel sum rule [see Eq. (\ref{eq:Sz_quantized}) and also Ref. [\onlinecite{Rouco19_YSR_and_ABS_in_SFS_junctions}]. We stress that these transitions are a generalization of the well-known ``0-$\pi$" transition occurring in atomic Shiba impurities \cite{Sakurai70, Franke11_Competition_of_Kondo_and_SC_in_molecules} or quantum dots  coupled to superconductors \cite{Bauer07_NRG_Anderson_model_in_BCS_superconductor, Deacon10_Tunneling_spectroscopy_of_ABS_in_SC_QDs, Lee14_Spin_resolved_ABS_in_SU_SE_nanostructures}. From this perspective, the difference with respect to atomic systems is that instead of a single transition, actually multiple transitions can occur due to the finite extension $L$ of the ``impurity" and the many ABS states with different symmetry which can eventually cross below $E_F$. Interestingly, we stress that the ocurrence of these quantum phase transitions can be tuned varying the length $L$. 

We now briefly address the effect of the Rashba spin-orbit interaction, which has been neglected in our work. As mentioned previously, this interaction was neglected to simplify the theoretical description of this (already quite complex and rich) problem. This interaction can drive the system into the topological superconductor class D \cite{Altland97_Symmetry_classes, Ryu10_Topological_classification}, hosting Majorana zero modes at the ends (see e.g., Ref. \onlinecite{Sau15_Bound_states_in_a_FM_wire_in_SC} for a related setup), and in that case we expect qualitative changes with respect to the results presented here. Consequently our results apply to experimental 
SE-SU-FMI systems where the spin-orbit energy term $E_\text{SOC}=\alpha_R^2 m^*/2$, with $\alpha_R$ the Rashba parameter, is negligible compared  to $\Delta$ and $h_0$. 

Finally, we consider the effect of disorder in this setup. This might be a relevant effect as a random disorder potential will eventually break the inversion symmetry of the model and might lift the predicted even-odd degeneracy in the limit $k_FL \gg 1$. However, we believe the energy-lifting effect might be weak in epitaxially-grown samples, where disorder is a relatively small effect.

\acknowledgments

This work was partially supported by CONICET under grant PIP 0792, UNLP under grant PID X497, and Agencia I+D+i under PICT 2017-2081, Argentina. AML is grateful to Liliana Arrachea for pointing out crucial bibliographic references.

%\bibliography{thispaper}


\end{document}